\documentclass{aa}
\usepackage{natbib}
\usepackage{amssymb}
\usepackage{amsmath}
\usepackage{graphicx}
\DeclareGraphicsExtensions{.eps, .ps, .eps.gz, ps.gz}
\usepackage{placeins}
\usepackage{subfigure}
\usepackage{color}
\usepackage[T1]{fontenc}

\def\Mpc{\, h^{-1} \, {\rm Mpc}}
\def\kpc{\, h^{-1} \, {\rm kpc}}
\def\Mo{\, h^{-1} \, {\rm M_{\odot}}}

\begin{document}
  
\title{
What determines large scale galaxy clustering: halo mass or local density?
}
\author
{Arnau Pujol \inst{1} \and 
Kai Hoffmann \inst{1} \and 
Noelia Jim\'{e}nez \inst{1,2} \and 
Enrique Gazta\~{n}aga\inst{1} \\
}
\institute{
Institut de Ci\`{e}ncies de l'Espai (ICE, IEEC/CSIC), E-08193 Bellaterra (Barcelona), Spain\\
\and
School of Physics \& Astronomy, University of St Andrews, North Haugh, St Andrews KY16 9SS, Scotland, UK
} 

\date{Received date / Accepted date}


\abstract{
Using a dark matter simulation we show how halo bias is determined by local density and not by halo mass. This is not totally surprising as, according to the peak-background split model, local matter density ($\bar \delta$) is the property that constrains bias at large scales.
Massive haloes have a high clustering because they reside in high density regions. Small haloes can be found in a wide range of environments  which determine their clustering amplitudes differently.  This contradicts the assumption of standard
Halo Occupation Distribution (HOD) models that the bias and occupation of haloes is determined solely by their mass.
We show that the bias of central galaxies from semi-analytic models of galaxy formation as a function
of luminosity and colour is therefore not correctly predicted by the standard HOD model.
Using $\bar \delta$ (of matter or galaxies) instead of halo mass the HOD model correctly predicts galaxy bias. 
These results indicate the need to include information about local density and not only mass in order to correctly apply HOD analysis in these galaxy samples.
This new model can be readily applied to observations and has the advantage that the galaxy density can be directly observed, in contrast with the dark matter halo mass.
}


\maketitle



\section{Introduction}
In the standard cosmological framework, the so-called ${\Lambda}$CDM
paradigm, galaxies form, evolve, and reside in dark matter haloes that
grow and assemble in a hierarchical way \citep{White1978}.
Therefore, it is assumed that the intrinsic properties and the evolution of these host haloes will
have an impact in the subsequent galaxy population inhabiting the haloes. The
unknown nature of the dark matter (and dark energy), could be unveiled
through the study of the baryonic observables (such as galaxies), once we properly understand
their co-evolution and this task, modelling the relation between the observed galaxy distribution and the underlying dark matter field, is a fundamental question of modern cosmology.

Given a cosmology, the halo mass function, the halo concentration and halo bias can be defined. This implies that the dark matter density field is not perfectly mapped by the distribution of the dark matter haloes, since they are biased tracers of it. In order to model the galaxy clustering statistics it is necessary to specify the
number and spatial distribution of galaxies within these dark matter haloes. One simple approach is to use the statistical halo distribution known as the standard 'Halo Occupation Distribution'   \citep[hereafter referred to as HOD,][]{Jing1998,Benson2000,Seljak2000,Scoccimarro2001,Cooray2002,Berlind2002}.  Its simplicity resides in the common assumption  that the derived occupation parameters and the physical properties of the galaxies are solely determined by
the mass of the halo in which they reside. Therefore, the probability that a halo of mass $m$ hosts $N_{gal}$ galaxies of a given property is given by the quantity  $P(N_{gal}|m)$.
 
The HOD has been proven a very powerful theoretical tool to constrain
both the galaxy-halo connection and the fundamental parameters in
cosmology \citep{Yang2003,Zehavi2005,Cooray2006,vdBosch2007,Zheng2007,Zehavi2011,Tinker2013}. 
By the use of the HOD model mock galaxy catalogues are constructed for studies of galaxy formation, as well as for the preparation and analysis of observational surveys \citep{Berlind2002,Zheng2007, BOSS2015,Carretero2015}.
 However, it has become clear that the clustering of dark
matter haloes depends on other properties besides mass \citep{Sheth2004,Gao2005,Wechsler2006,Gao2007,Jing2007,Wetzel2007, Falta2010,Lacerna2011,Lacerna2014,Pujol2014}. For example:
the construction of mock galaxy catalogues through N-body simulations has
shown that the amplitude of the two-point correlation function of
dark matter haloes with masses lower than $10^{13}\Mo$ on large scales
depends on the halo formation time \citep[e.g.,][]{Gao2005}.  Additionally, 
halo properties such as concentration, shape, halo spin, major merger rate, triaxiality, shape of the velocity ellipsoid, and velocity anisotropy, show correlations with other properties than  halo mass  \citep{Bett2007,Croton2007,Lacerna2012}.  Moreover, haloes of equal mass could have different galaxy occupation statistics, depending on their  environment \citep[e.g.,][]{Croft2012,Pujol2014}. Ignoring the effects of properties other than mass in the HOD modelling could distort the
 conclusions and interpretations of the observational results  \citep{Zentner2014}.  
 
To solve this issue, new studies point in the direction of taking into account these halo properties linked with the environment and their formation history and modify the standard HOD accordingly.  The relevant properties can be incorporated in the formulation of more flexible schemes of HOD used to produce mock catalogues comparable with observations.  Such  catalogues have recently been presented by \cite{Croton2007} and \cite{Masaki2013}, who introduced a rank ordering of galaxy
colours or Star Formation Rate (SFR). The authors point that for fixed halo mass, red galaxies are more clustered than blue galaxies. These and the 'age matching' models of \citet{Hearin2013,Hearin2014} have been shown to successfully reproduce
a number of observed signals such as the two-point clustering and
galaxy-galaxy lensing signal of SDSS galaxies (\citealt{Lacerna2014} and references therein). Recently, \cite{Hearin2016} presented the idea of decorated HOD, a new parametrization that allows the galaxy sample to be affected by a parametrized assembly bias. 

According to the peak-background split model \citep{Bardeen1986,ColeKaiser1989},
the fundamental quantity that describes halo bias is the density fluctuation of the dark matter field
 and not mass. Hence, we expect local density to constrain bias better than mass. 
In this paper we show the dependence of large scale halo bias on mass and local density (defined as the fluctuations of the
local background density $\bar \delta$), and confirm the peak-background split prediction that density is the property that constrains bias. 

The aim of our study is to analyse how well mass and local density
determine galaxy bias and HOD.  For this, we present a bias reconstruction method that tests how well the standard HOD assumptions (that halo bias and occupation depend only on mass) are able to correctly predict galaxy bias. To do this, we compare the measured galaxy bias with the one reconstructed from the measurements of halo bias and HOD. This method was already implemented in \cite{Pujol2014}, where we showed this test for luminosity dependent galaxy bias. In this paper, we also use the reconstruction method to test how well local density is able to predict galaxy bias from halo bias and HOD. We find that mass is not able to predict galaxy bias for colour selected samples. In these cases, local density makes a much better prediction.

Our analysis is based on haloes from the Millennium simulation \citep{Springel2005} and the Semi-Analytical Model (SAM) of galaxy formation of \citet{Guo2011}. The SAM  populate  haloes with galaxies that are evolved and followed in time
inside the complex structure of merger-trees. The baryonic processes
included are laws for metal dependent gas cooling, reionization, star
formation, gas accretion, merging, disk instabilities, AGN and
supernovae feedback, ram pressure stripping and dust extinction, among others
\citep[e.g.,][]{Baugh2006,Jimenez11,Gargiulo2015}.  Because of these processes, the
resulting galaxy population produced by SAMs is sensitive to the environment and evolution of dark matter haloes. Depending on the galaxy selection, the information from halo mass alone could be insufficiently correlated with the clustering of galaxies. In these cases, local density allows to determine galaxy bias better than halo mass. Thus, we propose to use the information from local density to improve the HOD analyses on surveys and theory. 


The paper is organized as follows. In section \ref{sec:methodology}
we describe the data used and the methodology for our measurements of clustering, bias
and for our HOD reconstructions. The results are shown
in section \ref{sec:results}, and we summarize the conclusions of the
paper in section \ref{sec:conclusions}.

\section{Methodology}\label{sec:methodology}

\subsection{Simulation data}\label{sec:simulation}

In this study we use the data from the Millennium Simulation \footnote{\url{http://www.mpa-garching.mpg.de/millennium/}} \citep{Springel2005}, an N-body simulation generated using the \textsc{GADGET-2} code \citep{Springel2005b}. The simulation corresponds to a $\Lambda$-CDM cosmology with the following parameters: $\Omega_m = 0.25$, $\Omega_b = 0.045$, $h = 0.73$,  $\Omega_\Lambda = 0.75$, $n = 1$ and $\sigma_8 = 0.9$. It contains $2160^3$ particles in a comoving box with a side length of  $500\Mpc$. The resolution corresponds to a particle mass of $8.6 \times 10^8 \Mo$ and a spatial resolution of $5\kpc$. The cosmological model is based on WMAP-1 \citep{Spergel2003} the 2-Degree Fields Galaxy Redshift Survey (2dFGRS) data \citep{Cole2005}. The initial conditions have been calculated using CMBFAST \citep{Seljak1996}. 
We use the comoving output at $z=0$.

The haloes are identified using the Friends-of-Friends (FOF) algorithm, using a linking length of $0.2$ times the mean particle separation, and discarding all haloes with less than $20$ particles. In our analysis we define the halo mass from the total number of particles belonging to the FOFs. 

Galaxy catalogues of several Semi-Analytical Models (SAM) are available in the public database of the simulation. For this analysis we use four models \citep{Bower2006,DeLucia2007,Font2008,Guo2011}. Although the results are different for each model, they all show the same behaviours and the conclusions of our study do not depend on to SAM used. For this, we show the results from \cite{Guo2011} in section \ref{sec:results}, and we show in Appendix \ref{sec:appendix} a comparison of the rest of the models. 

\subsection{Clustering and bias} \label{sec:clustering_and_bias}

Spatial fluctuations of the matter or tracer density $\rho$ are defined as normalised deviations from the mean density $\bar \rho$ at the position $\bf{r}$, i.e.

\begin{equation} 
	\delta(\bf r) \equiv \frac{\rho(\bf r) - \bar{\rho}}{\bar{\rho}}.
	\label{eq:dc_definition}
\end{equation}
Note that in this article we will often refer to these density fluctuations as density for simplicity. We measure $\delta(\bf r)$ by dividing the simulation into cubical grid cells with side lengths of $500/64 \sim  8\Mpc$ 
and assigning a $\delta$ to each cell. We then measure the two-point correlation function as
\begin{equation}
	\xi_{AB}(r) \equiv \langle \delta_A({\bf r_1}) \delta_B({\bf r_2}) \rangle,
	\label{eq:def_xi}
\end{equation}
which is a function of the scale $r \equiv |{\bf r_2}-{\bf r_1}|$. The average
$\langle \ldots \rangle$ is taken over all pairs of $\delta$ in the analysed volume, independently
of their orientation. The indices $A$ and $B$ refer to the fluctuations of different density tracers (here haloes or galaxies,
which we generically call $\delta_{G}$)  or to those of the matter density, which we denote as
$\delta_m$. Hence, $A=B=m$ denotes the matter auto-correlation ($\xi_{mm}$), while
$\xi_{Gm}$ is the cross-correlation between the tracer $G$ and matter $m$.
We then estimate the bias by the ratio:
\begin{equation}
    b(r)  \equiv \frac{\xi_{Gm}(r)}{\xi_{mm}(r)}.
	\label{eq:b1_xi}
\end{equation}
 At large scales ($r\gtrsim20 \Mpc$), where $\xi<1$, this ratio is well described by a constant, which is in good agreement with the linear bias
and tends to the value $\delta_G \simeq b_1 \delta_m$, 
  (e.g. see \citealt{Bel2015}).
We estimate this linear bias $b_1$, which we just call $b$ from now on,
by fitting $b(r)$ with a constant in the scale range 
of $20\Mpc < r < 30\Mpc$.  For larger scales, the measurements start to be noisy 
due to the size of the simulation.
The covariance of these measurements are derived by jack-knifing
\citep[e.g.][]{Norberg2009}, using $64$ cubical subvolumes.

The value of $b$ depends on how the tracers are selected. In galaxy surveys
the limits of the observed galaxy luminosities roughly correspond to a selection
by host halo mass.
The mass dependence of the bias can be predicted from the peak-background
split  model (see \citealt{Bardeen1986, ColeKaiser1989, MoWhite1996}
and \citealt{Hoffmann2015} for a recent validation of the predictions)
where the matter density field is described by the superposition
of small scale fluctuations (peaks) $\delta_s$ with large-scale fluctuations of
the background matter density $\bar \delta$ around each peak
(i.e. $\delta = \delta_s +  \bar \delta$).
 In this model background fluctuations can lift the peak-heights above
a critical density $\delta_{crit}$ at which they collapse to haloes as illustrated by Fig.\ref{fig:PBS}.
The mass of these haloes then
corresponds to the peak-heights. Consequently massive haloes tend to reside in environments
where the background density is high, while low mass haloes can reside in a broader range
of environmental densities (we show this effect with haloes from the Millennium
simulation in Section \ref{sec:hal_bias}). Massive haloes are therefore more strongly clustered
than low mass haloes, which causes a mass dependence of the bias parameter.

\subsection{Local background density}

The bias of a given halo sample can also depend on additional halo 
properties besides the halo mass, such as the concentration of the mass density 
profile or the properties of the galaxies hosted by the halo \citep[see e.g.][]{Gao2005, Falta2010, Hearin2015}.
These additional properties can be related to the merging history of
haloes and lead to the so-called assembly bias. Furthermore, the bias of a given halo
sample is affected by the tidal forces from the large-scale 
environment, which is known as non-local bias \citep[see][]{Chan12, Baldauf12, Bel2015}.

\begin{figure}
    \centering
        \vspace{-30pt}
    \includegraphics[scale=.30, angle =270]{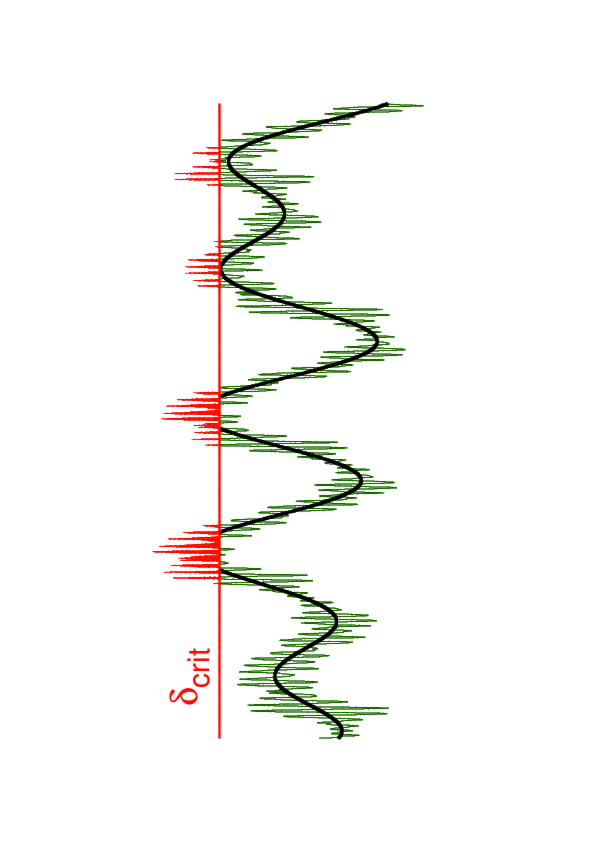}
        \vspace{-50pt}
    \caption{Illustration of peak-background split  
    model,  $\delta = \delta_s +  \bar \delta$, showing in red the location of haloes
that form above some  critical value $\delta_{crit}$.  The clustering of those haloes
is stronger than the rest (in green). But note how once the local density $\bar \delta$ (black curve) is fixed at
$\delta_{crit}$,  the large scale clustering  of haloes (red) does not depend on the peak-hight (halo mass).}
    \label{fig:PBS}
\end{figure}

These effects can lead to wrong predictions of models which rely on the assumption that
the mass of a halo sample completely determines the bias, such as self-calibration techniques
(e.g. \citet{Wu08}) or HOD models \citep{Zentner2014, Pujol2014}.
A way to circumvent problems for the HOD model is to use a halo property
different from the mass, which completely determines the bias of a given halo population.
The peak-background split argument described in Subsection \ref{sec:clustering_and_bias} suggests that the clustering of peaks (which
correspond to tracers such as haloes) is determined by fluctuations of the
large-scale background density $\bar \delta$. Hence, for fixed background densities the bias
should be independent of the tracer properties (see Fig.\ref{fig:PBS} for illustration).

\begin{figure}
    \centering
    \includegraphics[scale=.45, angle =270]{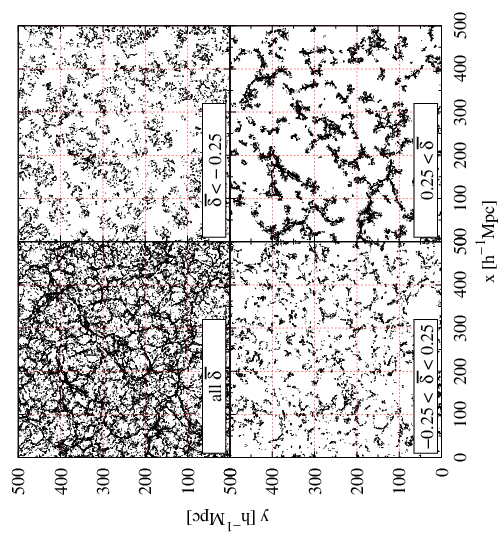}
    \caption{Distribution of haloes with different dark matter local densities $\bar \delta$ in a $2 \Mpc$ slice of the Millennium simulation.
    The top left panel shows all haloes, while the top right shows haloes in under-densities. The bottom left shows haloes with mean background densities while the bottom right shows haloes in over-dense regions.}
    \label{fig:xz_deltamassbins}
\end{figure}

For obtaining a visual impression of this argument we show in Fig. \ref{fig:xz_deltamassbins}
the spatial distribution of haloes which reside in regions with different densities, defined by the value of
$\bar \delta$ estimated around each halo, smoothed on cubical cells of side $l = 14 \Mpc$ as explained below.
 By comparing the four panels of the figure one can see that
 the large-scale clustering and hence the bias strongly changes
with the local dark matter back-ground density $\bar \delta$ (as we will show in Section \ref{sec:hod_dm}). 
On the other hand the large-scale clustering is nearly independent of  halo mass
(not shown in this figure), when haloes are selected by $\bar \delta$. Note that a weak mass dependence of the
bias for fixed $\bar \delta$ can be expected from  the aforementioned non-local bias.

When determining $\bar \delta$ around a given halo we face the problem that the dark
matter density distribution of the Millennium Simulation is publicly available only as a
$500 / 256 \simeq 2 \Mpc$ grid. We therefore assign the density in a cubical
volume around each halo, with length $l \simeq 6, 14$ and $26 \Mpc$, which have the same
volume as a sphere of radius $R = 3.72, 8.68$ and $16.13 \Mpc$ respectively. The position
of these volumes has a $2 \Mpc$ inaccuracy, which results from the grid cell size.
In the presentation of our results we will focus on the intermediate scale $R=8.68\Mpc$,
while this choice does not affect our conclusion as results for the other scales
are similar.

\begin{figure}
    \centering
    \includegraphics[scale=.43]{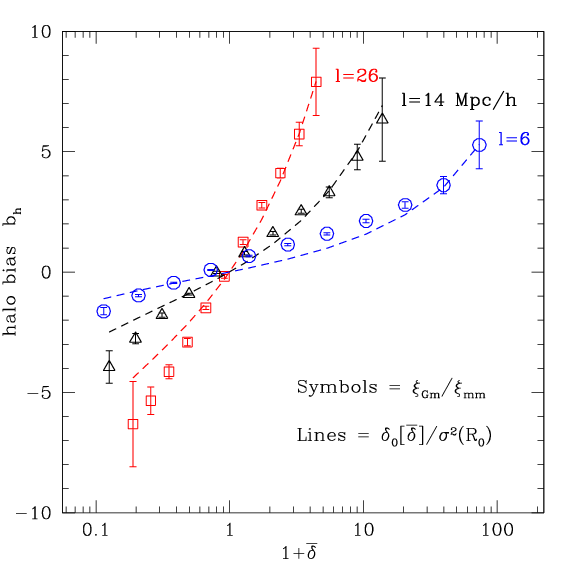}
    \caption{Halo bias $b_h$ as a function of mass local density $\bar{\delta}$
    in cubical volumes with side length $l=26 \Mpc$ (red), $l = 14\Mpc $ (black) and $l = 6\Mpc$ (blue),
     which have the same volume as a sphere of radius $R = 3.72, 8.68$ and $16.13 \Mpc$ respectively.
      The corresponding dashed line shows a simple prediction $b \simeq \delta_0/\sigma_0^2$
    where $\bar{\delta}_0$ is the linear density corresponding to $\bar{\delta}$ (according to the spherical collapse model)
    and $\sigma_0^2$ is the linear variance at Lagrangian scale $R_0=R (1+\bar{\delta})^{1/3}$.}
    \label{fig:bias_delta_theory}
\end{figure}

The linear halo bias $b_h$ is shown as a function of $\bar \delta$ in Fig. \ref{fig:bias_delta_theory}
using haloes of all masses. The bias increases with the density of the environment and becomes negative in
underdense regions. Note that we find negative bias because we study the halo-matter cross-correlation,
while in the case of the auto-correlation the bias would remain positive.
The dependence of $b_h$ on $\bar \delta$ becomes weaker when $\bar \delta$ is defined at smaller scales $R$.
Note that the clustering of galaxy with different environmental densities has been studied in data by \citet{AbbasSheth2007}.

Since the large-scale clustering of the tracers corresponds  to the clustering of the background density
fluctuations in which they reside we can compare our measurements to predictions for the clustering of 
the excursion set model \citep{Sheth1998}. We use equation $5$ from \cite{AbbasSheth2007}:

\begin{equation} 
	b \simeq \frac{\bar \delta_0}{\sigma_0^2},
	  \label{eq:bias_delta_theory}
\end{equation}
where ${\bar \delta_0}$ is the initial linear overdensity and $\sigma_0$ is the corresponding linear variance on the Lagrangian scale $R_0 = R (1 + \bar \delta)^{1/3}$. To estimate ${\bar \delta_0}$ we use the spherical collapse model to map the measured non-linear local density into the initial density \citep[see also][]{FG98}. These predictions are in general in a good agreement with the measurements and do not include any free parameters.
Note  that some of the differences at small $\bar \delta$ could come from the fact that
we measure the bias in bins of $\bar \delta$, while the prediction are for threshold values of ${\bar \delta_0}$, so we expect
them to be systematically higher. This can be easily corrected, but is beyond the scope of this paper.

In the next subsection we will directly use these bias  measurements  $b_h(\bar{\delta})$ to study the bias predictions from the standard HOD assumptions, and to  present a new HOD test, based on local density $\bar{\delta}$, which is less affected by assembly bias.

\begin{figure*}
    \centering
    \includegraphics[scale=.43]{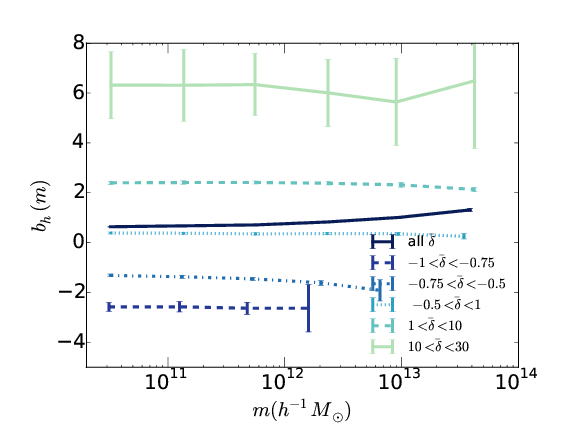}
        \includegraphics[scale=.43]{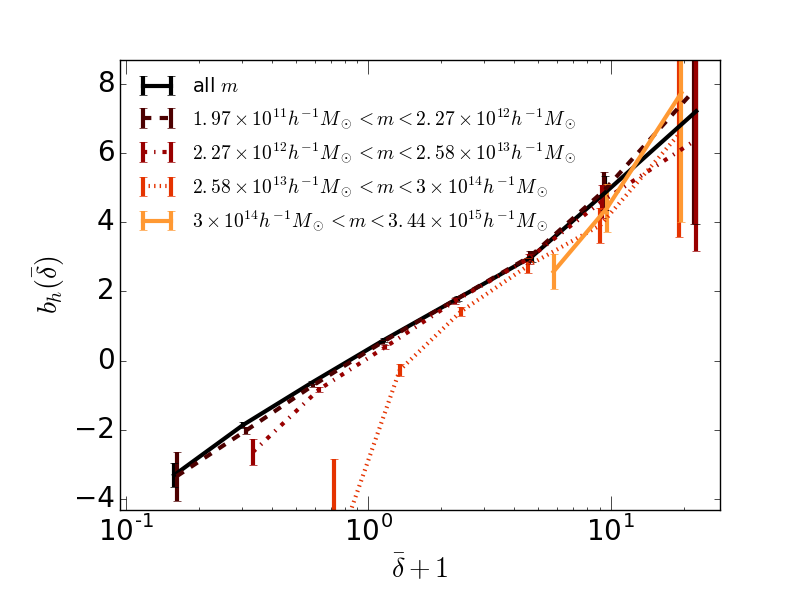}
    \caption{Left: halo bias as a function of mass. Each line corresponds to haloes in a fixed density fluctuations $\bar \delta$. Right: halo bias as a function of the local density $\bar \delta$. Each line corresponds to a
    different halo mass bin. In both panels $\bar \delta$ is defined as the Eulerian density fluctuation around a cubic box of  $14\Mpc$ of side.}
    \label{fig:bh_mass_delta}
\end{figure*}


\subsection{Mass and density HOD bias reconstruction}
\label{sec:hod_model}

The standard HOD model is based on the assumption that halo bias
is determined solely by the halo mass $m$ of the sample. We will therefore
refer to it as the mHOD model in the following.
In this mHOD model the bias of galaxies  ($b_g$) which are selected by an arbitrary
property $P$,  are reconstructed from the halo bias as a function of mass $b_h(m)$  and
 the mean number of galaxies with property $P$ per halo with mass $m$, $\langle N_g(m | P) \rangle$ as

\begin{equation}
   	b^m_{rec}(P) = \frac{\int dm~b_h(m)~ \langle N_g(m | P) \rangle}{\int dm~ \langle N_g(m | P) \rangle }.
	\label{eq:bg_rec_m}
\end{equation} 
This equation can be seen as a weighted average of $b_h(m)$, where the weight is given by $\langle N_g(m | P) \rangle$. 
The mHOD model provides a way to infer the average mass of host haloes
in which a given galaxy population is residing by varying $\langle N_g(m | P) \rangle$ in order to
reproduce the observed bias, i.e. $b_{rec}^m(P) = b_g(P)$. However, it relies on the assumption
that the bias of a galaxy population selected by the property $P$
is completely determined by their host halo mass, which might not be correct as discussed
in Section \ref{sec:clustering_and_bias}. Cosmological N-body
simulations allow us to test the mHOD model, as
we can measure $b_h(m)$ and $\langle N_g(m | P) \rangle$ and compare the predicted $b_{rec}^m(P)$
to measurements of $b_g(P)$. This test was explored by \cite{Pujol2014}
and will hereafter be referred to as the bias reconstruction method. The test revealed that the
mHOD model fails to predict the bias correctly in the low halo mass range
but it works with ~5-10\% bias accuracy in the high mass end.

In the previous subsection we considered to redefine the
HOD model using the local background density $\bar \delta$ around a given halo instead of the
halo mass. This approach has the advantage that $\bar \delta$ is expected to determine
the bias of a given halo or galaxy population better than the mass
(we study this in Section \ref{sec:results}). Furthermore, $\bar \delta$ is
well defined, whereas the halo mass can vary significantly, depending on the definition.
In this paper we call the density HOD model (hereafter referred to as dHOD model) to the analogous of the mHOD model using density instead of mass. The corresponding bias reconstruction is obtained simply by
replacing the halo mass in equation (\ref{eq:bg_rec_m}) by $\bar \delta$,

\begin{equation}
   	b^{\bar \delta}_{rec}(P) = \frac{\int d{\bar \delta}~b_h(\bar \delta)~ \langle N_g({\bar \delta} |P) \rangle}{\int d\bar\delta~ \langle N_g({\bar \delta} | P) \rangle},
	\label{eq:bg_rec_delta}
\end{equation} 
In the next Section we will test how well our new dHOD model predicts the bias compared
to the standard mHOD model.


\begin{figure}
    \centering
    \includegraphics[scale=.4]{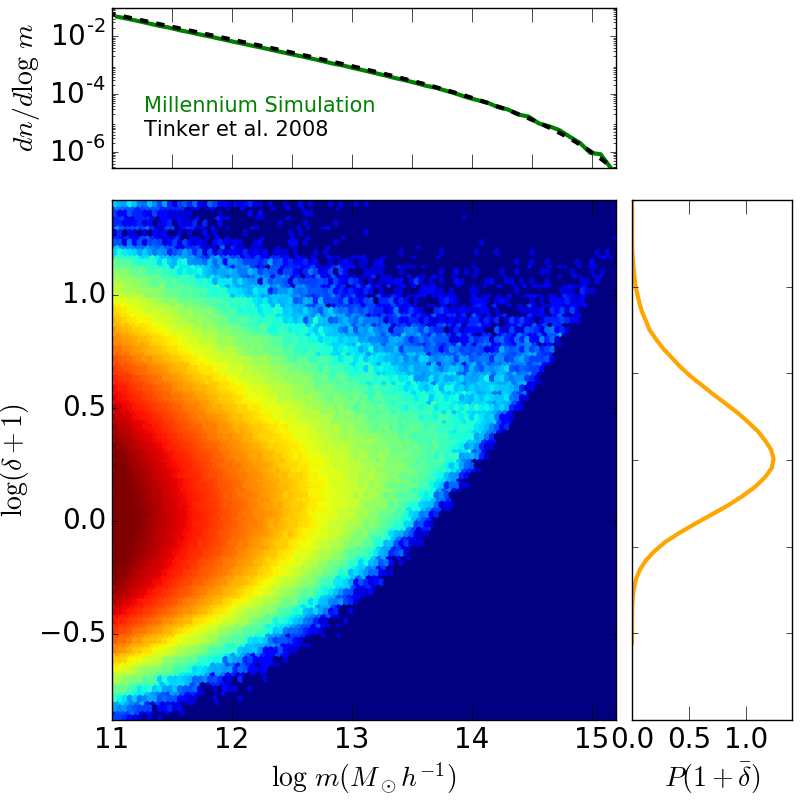}
    \caption{Distribution of haloes in local density $\bar \delta$ and mass $m$. The colours show the number
    of haloes that are in a density $\bar \delta$ and have a mass $m$. Right panel shows the
    $\bar \delta$ distribution from the contribution of all the haloes, $P(1+\bar \delta)$. Top panel shows the Halo Mass Function, so the
    contribution of all the haloes in the mass distribution.}
    \label{fig:mass_delta}
\end{figure}

\section{Results}\label{sec:results}

\subsection{Halo bias}\label{sec:hal_bias}

In this Subsection we analyse the dependence of halo bias on halo mass and local background matter density fluctuations $\bar \delta$ 
(smoothed with a cubical top hat filter with a side length of $l = 14 \Mpc$ side,
which has the same volume as a sphere of radius $R = 8.68\Mpc$) in the environment of each halo. The measurement of the latter is
described in Section \ref{sec:methodology}. Similar results are found for other smoothing scales.

In the left panel of Fig. \ref{fig:bh_mass_delta} we show the halo bias, measured via equation (\ref{eq:b1_xi}) (hereafter referred to as $b_h$)
as a function of the halo mass $m$ for haloes within different $\bar \delta$ ranges. The black line corresponds to $b_h(m)$ for
all haloes, selected independently of $\bar \delta$.
This latter measurement is consistent with the theoretical model of \citealt{Tinker2010} (see Fig. 3 of \citealt{Pujol2014}).
We can see that $b_h(m)$ does not change significantly (less than a $10\%$) with halo mass when $\bar \delta$ is fixed, while it changes significantly when all the haloes are included (around a factor of $2$ in the range of masses shown here).
In the right panel we show the same analysis from a different point of view. Here we present $b_h(\bar \delta)$
for different $m$ bins. Each line corresponds to a range in $m$, while the black solid line 
shows $b_h(\bar{\delta})$ for all the haloes. We can see a strong dependence of $b_h$ on $\bar \delta$, and that $b_h(\bar \delta)$
depends weakly on $m$.

These measurements demonstrate the argument from the peak-background split that local density is the property that constrains clustering. We show that the bias of a given halo sample is well constrained when the haloes are solely selected by the local density in their environment $\bar \delta$, almost independently of the halo mass.

If $\bar \delta$ is left as a free parameter, a mass dependence of the
bias arises from the fact that high mass haloes tend to reside in high density regions and low mass haloes
in regions with lower density. This tendency can be seen in the central panel of Fig. \ref{fig:mass_delta}, where we show the
mass versus the local density $\bar \delta$ for each halo in the simulation. The colours describe the number of haloes with the corresponding mass and density. By integrating this distribution over
$\bar \delta$ we derive the Halo Mass Function (hereafter referred to as HMF), $\langle N_h(m,\bar \delta)/(V\log(m))\rangle_{\bar \delta}$,
(where $V$ is the simulation volume and $N_h$ the absolute number of haloes). The HMF
is well described by the \cite{Tinker2008} model, as shown in the top panel of Fig. \ref{fig:mass_delta}.
By integrating over the mass $m$ we obtain the Probability Distribution Function (hereafter referred to as PDF) of $\bar \delta$, or
$\langle N_h(\bar \delta,m)/n(m)\rangle_m$, which is roughly log-normal, as shown in the right panel of
Fig. \ref{fig:mass_delta}.

\begin{figure}
    \centering
    \includegraphics[scale=.43]{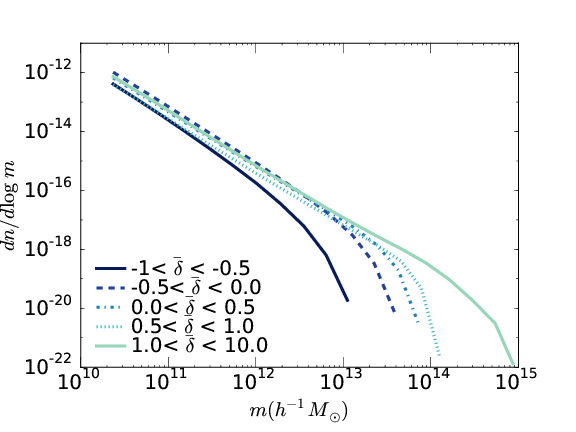}
    \includegraphics[scale=.43]{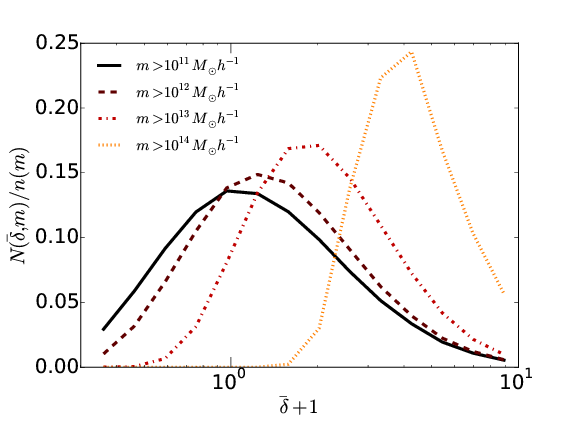}
    \caption{Distributions of haloes in $m$ and $\bar \delta$. Top panel shows the HMF for haloes in different densities.
    Each colour represents haloes of a range in $\bar \delta$. The bottom panel shows the PDF of $\bar \delta$ of haloes of
    different masses. Different colours represent haloes of different $m$ ranges.}
    \label{fig:hod_delta_mass}
\end{figure}

The aforementioned tendency that high mass haloes tend to reside in high density
regions and low mass in regions with lower density can be seen more clearly in Fig. \ref{fig:hod_delta_mass}.
In the top panel of this figure we show the HMF of haloes with different background densities $\bar \delta$.
We find that the fraction of massive haloes decreases in low densities, while fraction of low mass haloes
is similar. The same effect can be seen in the bottom panel of this figure, where we show the PDF of $\bar \delta$
for haloes in different mass bins. We find that that massive haloes (i.e $m>10^{14}\Mo$) 
reside almost exclusively in very dense regions and are unlikely to be found in regions with low density, while
low mass haloes can be found in a wide range of densities, with preference to average values of $\bar \delta = 0$.

The dependence of the $b_h(m)$ measurements on the local halo background density $\bar \delta$ is in agreement
with \cite{AbbasSheth2007}. This result can be expected, since $\bar \delta$ is defined from larger scales than $m$, and larger scales are expected to determine better bias.
The standard HOD model assumes that halo bias is determined solely by the halo mass, but the local density contains additional information that constraints bias. 
We therefore test the
accuracy of mHOD reconstructions in the next subsection and compare it to reconstructions from our new dHOD model from
equation (\ref{eq:bg_rec_delta}), which is based on the assumption that the bias is determined solely by $\bar \delta$,
as discussed in Section \ref{sec:methodology} and suggested by the results of this Subsection.

\begin{figure}
    \centering
    \includegraphics[scale=.43]{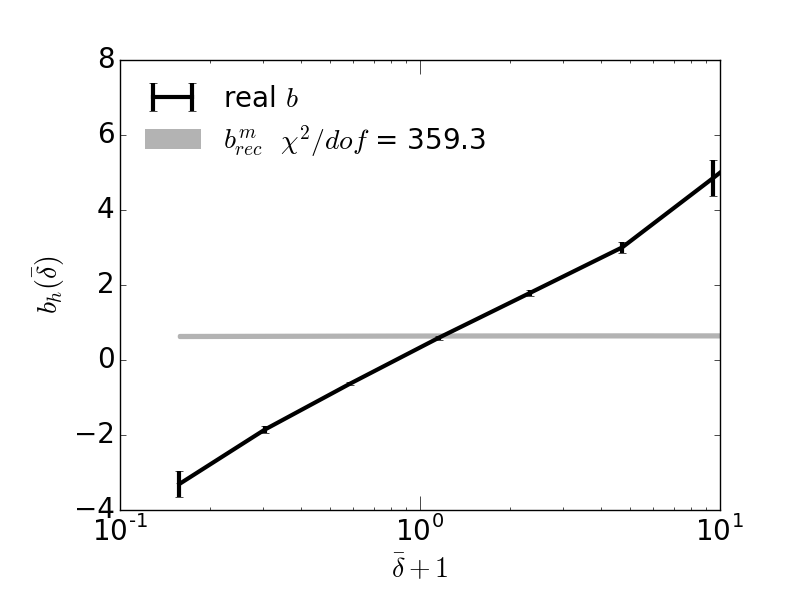}
    \includegraphics[scale=.43]{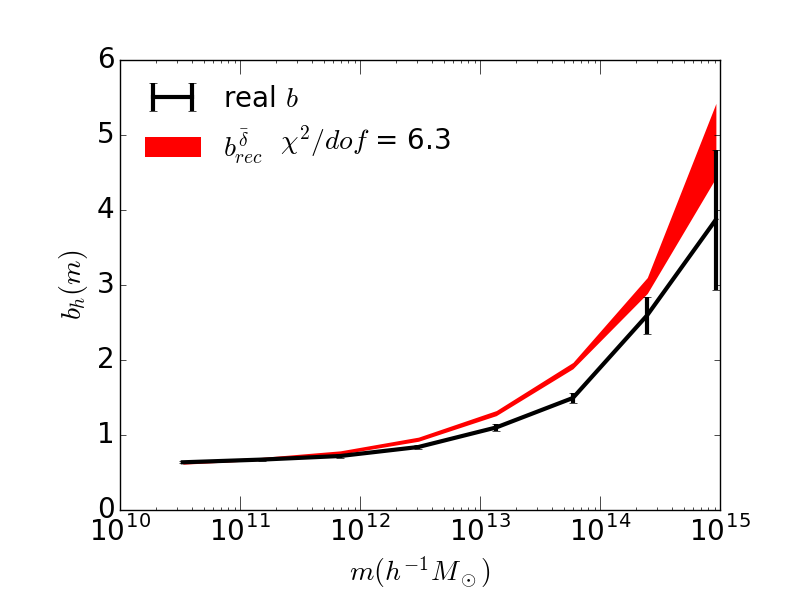}
    \caption{Halo bias reconstructions compared to the measured bias from the simulation. The
    solid black lines show the measurement of $b_h$ from the simulation. The coloured regions show the $1\sigma$
    interval of the reconstructions. Top panel shows $b_{rec}^m(\bar \delta)$ (in grey)
    from $b_h(m)$, while bottom panel shows  $b_{rec}^{\bar \delta}(m)$ (in red) from $b_h(\bar \delta)$.}
    \label{fig:rec_bias}
\end{figure}

\subsection{HOD tests using Mass and Density}\label{sec:hod_dm}

In this subsection we study how well mass $m$ and local density $\bar{\delta}$ determine the linear 
 bias of a given halo sample.
 We do this by testing how well the mHOD model can predict the halo bias as a function of background density $\bar \delta$
and how well the dHOD model can predict the halo bias as a function of halo mass $m$.
The reconstruction for $b_h(P=\bar \delta)$ from the mHOD model is derived by averaging the $b_h(m)$ measurement
using the measured $\langle N_h(m | \bar \delta)\rangle$ as weight, as given by equation (\ref{eq:bg_rec_m}). This reconstruction
is then compared with measurements of $b_h(\bar \delta)$, derived from the two-point correlations via equation (\ref{eq:b1_xi}).
The dHOD reconstructions for $b_h(P=m )$ is tested in an analogous way using $b_h(\bar \delta)$ and $\langle N_h(\bar \delta | m) \rangle$
measurements in combination with equation (\ref{eq:bg_rec_delta}).

We can see in the top panel of Fig. \ref{fig:hod_delta_mass} that the HMF
depends on $\bar \delta$ only in the high mass end, while it is similar for different  $\bar \delta$
in the low mass end.
Since the low mass end of the HMF dominates the integral of the of the mHOD bias reconstruction 
$b_{rec}^m(\bar \delta)$ from equation (\ref{eq:bg_rec_m}), we do not expect
$b_{rec}^m(\bar \delta)$ to be strongly dependent on $\bar \delta$. The PDF
of $\bar \delta$ in the bottom panel of Fig. \ref{fig:hod_delta_mass} shows that haloes of
different $m$ are differently distributed in $\bar \delta$. Therefore we expect a strong
dependence of $b_{rec}^{\bar \delta}(m)$  from equation (\ref{eq:bg_rec_delta}) on $m$.

\begin{figure}
        \centering
        \includegraphics[scale=.43]{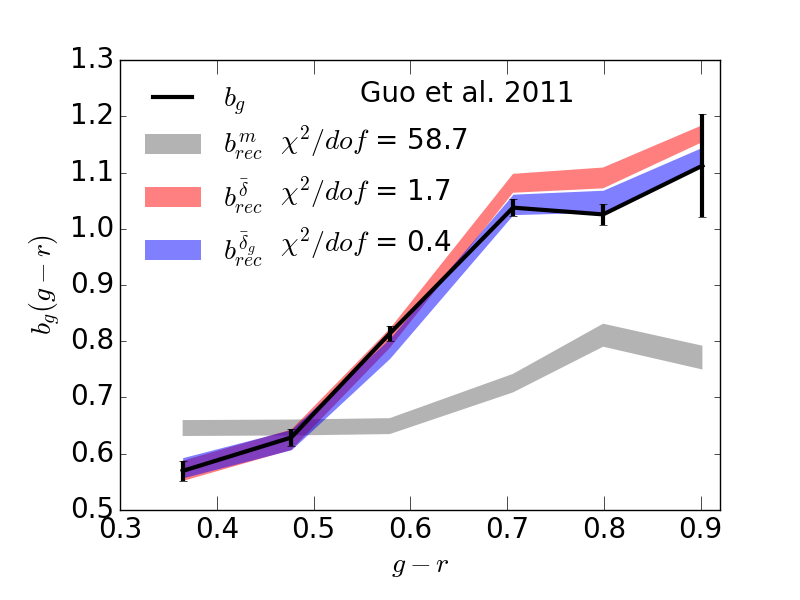}
        \includegraphics[scale=.43]{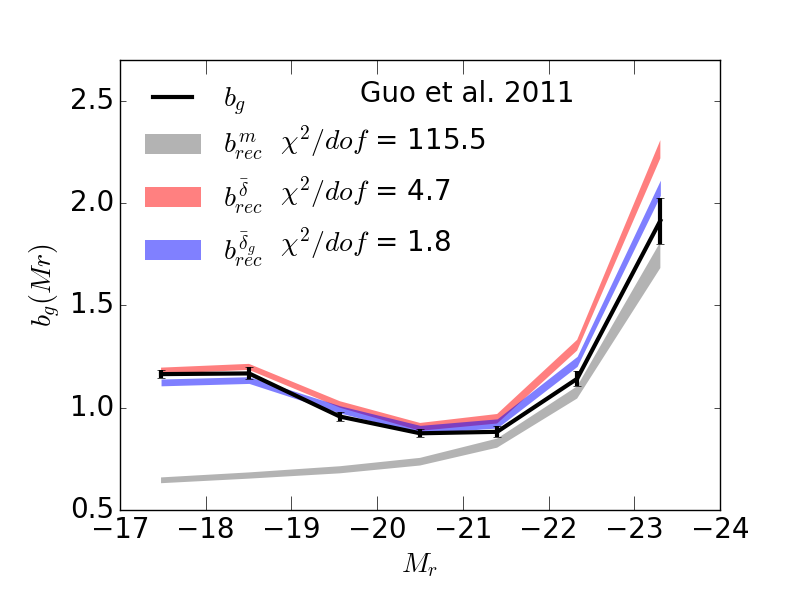}
        \caption{Galaxy bias compared to different reconstructions. The black solid lines show the measurements of
       $b_g$ from the simulation. The coloured regions show the $1\sigma$ level of the different reconstructions of $b_g$ from
        $b_h(m)$ (so $b_{rec}^m$, in grey), from $b_h(\bar \delta)$ (so $b_{rec}^{\bar\delta}$, in red) and from $b_h(\bar \delta_g)$ (so $b_{rec}^{\bar\delta_g}$, in blue).
        The top panel $b_g$ is for central galaxies as a function of colour $g-r$, and in the bottom panel $b_g$
        is displayed as a function of absolute $r$-band magnitude for red central galaxies (red defined as $g-r > 0.6$).}
        \label{fig:recs_mass}
\end{figure}

The bias reconstructions $b_{rec}^m(\bar \delta)$ and $b_{rec}^{\bar \delta}(m)$ are compared
with the measured bias $b_h(\bar \delta)$ and $b_h(m)$ in Figure \ref{fig:rec_bias}. In the top panel of Figure \ref{fig:rec_bias}, we show the measurement
of $b_h(\bar \delta)$ as solid black line. The grey region corresponds to $b_{rec}^m(\bar \delta)$ at the $1 \sigma$ level. The bottom panel shows $b_h(m)$, in solid black, and $b_{rec}^{\bar\delta} (m)$, coloured in red. The red region corresponds to the $1 \sigma$ errors of the reconstructions, derived from the jack-knife samples.
As expected from our considerations above, the $b_{rec}^m(\bar \delta)$ reconstruction does not show a significant $\bar \delta$ dependence and therefore it
deviates from the measured $b_h(\bar \delta)$.  This finding means that, once $\bar \delta$ is fixed,
the halo mass (or peak height) does not contain additional information about the large-clustering,
which already became apparent in the weak mass dependence of the bias of haloes with fixed $\bar \delta$, shown in Fig. \ref{fig:bh_mass_delta}. Hence the clustering of a given halo sample selected by $\bar \delta$ 
cannot be reconstructed via the standard mHOD model.

The $b_h(m)$ dHOD reconstruction, $b_{rec}^{\bar \delta}(m)$, shown in the bottom panel of
Figure \ref{fig:rec_bias}, is in much better agreement with the measurement. This finding demonstrates
that the bias is well determined by $\bar \delta$. The over prediction of $b_h$ by the dHOD model
at high halo masses results from the fact that $b_h$ is not completely independent of the mass at fixed
$\bar \delta$,  as we see in the left panel of Fig \ref{fig:bh_mass_delta}.

\subsection{HOD modelling of galaxy bias}

The results presented in the previous subsection confirm the peak-background split model and have important implications on
HOD modelling of galaxy clustering as galaxy properties are not only determined
by the mass of its host halo, but also by its interaction with the environment.

Assuming
that bias only depends on the halo mass can lead to a misinterpretation of HOD predictions (e.g. the fraction
of red satellites) and wrong predictions of galaxy bias as a function of galaxy
properties, such as colour or luminosity \citep{Pujol2014,Zentner2014}. In these cases, it
can be worthwhile to use the local density for HOD bias predictions, since we have
shown that it determines the bias better than halo mass.

In Fig. \ref{fig:recs_mass} we show the comparison of the different HOD reconstruction
methods in two different samples of central galaxies from the \cite{Guo2011} SAM. 
 In this analysis we focus on central galaxies, since their properties are more correlated with 
 the halo properties than satellite or orphans galaxies, and the implications of the local density dependence   of halo bias are more directly connected to this population.

The top panel shows the bias of central galaxies as a function of the colour index $g-r = M_g- M_r$,
where $M_g$ and $M_r$ are the absolute magnitudes in the SDSS $g$- and $r$-band,
taking into account dust extinction. The bias values below unity result from the fact that we did not apply a faint magnitude cut to the samples.
They are therefore dominated by dim galaxies with low clustering.
The bottom panel shows the bias of red central galaxies ($g-r>0.6$) versus $M_r$.
The solid black line shows the galaxy bias $b_g$ measured from the two-point correlation via equation
(\ref{eq:b1_xi}). The grey regions show the $1\sigma$ interval of the mHOD reconstruction for $b_g$
from equation (\ref{eq:bg_rec_m}), that indicates how well $m$ determines bias for these galaxy
samples. The red regions show the $1\sigma$ interval of the dHOD reconstruction of $b_g$
from equation (\ref{eq:bg_rec_delta}), that reflects how well $\bar \delta$ determines bias
for these galaxies. 

We can clearly see that the dHOD reconstruction is much closer to the measured bias 
than reconstruction from the mHOD model. This finding indicates that
 $\bar \delta$ determines $b_g$ much better than $m$.  
This is related to the formation times of the haloes and their assembly history. In terms of galaxy formation,  SAMs show that  galaxy colours are affected by the merging events (\citealt{Jimenez11} and references therein) which occur more often in high density environments.
We conclude that, when galaxy properties are determined by the halo local density, the standard mHOD
model can fail in predicting the bias as a function of that property. In this case HOD bias
predictions based on the more fundamental halo property $\bar \delta$, i.e. the dHOD model,
delivers more accurate bias predictions. This effect is not always as important as in
Fig. \ref{fig:recs_mass}, as we can see from \cite{Pujol2014}. In this work, $b_g$ is shown as a function of absolute magnitude, and since magnitude and mass are well related for these galaxies,  the mHOD reconstruction works well.

In real galaxy catalogues it is very difficult to measure the mean matter density, $\bar \delta$, that we are using to define environment. Instead we can easily measure $\bar \delta_g$, the density fluctuations of galaxies. In Fig. \ref{fig:recs_mass}
we also show the bias reconstruction using $\bar \delta_g$ instead of $\bar \delta$, so $b_{rec}^{\bar \delta_g}$, represented as blue regions. We measure $\bar \delta_g$ from galaxies of $M_r < -19$, and we clearly see that the result is  equivalent to that of using $\bar \delta$, in fact it is even better. This means that $\bar \delta_g$ determines bias in a similar way or better than $\bar \delta$. 
This result can be expected from the fact that the large scale fluctuations $\bar \delta$ and $\bar \delta_g$ are simply related to each
other by the linear bias of the background galaxy sample. Also note that  $\bar \delta_g$  is more closely related to the galaxy distribution
than  $\bar \delta$ (the relation between $\bar \delta_g$ and $\bar \delta$ might have stochasticity, for example) so it is not surprising that  $\bar \delta_g$ gives a slightly better reconstruction than $\bar \delta$.

It is important to mention that different galaxy formation models generally present different galaxy distributions. Because of this, the relation between galaxy bias and halo mass or local density can be different depending on the implemented galaxy formation model. For galaxy formation models based on the HOD, constructed using only the halo mass to define the galaxy populations, we would expect by construction a better agreement between $b_g$ and $b_{rec}^{m}$. However, these HOD models would be then totally insensitive to assembly bias effects. This is not the case of SAMs, since these models are constructed from the halo merger trees. This study has been done using 4 different SAMs in order to see how much our conclusions depend on different implementations of SAMs. We have seen that all SAMs present the same behaviour, and we show the comparison of these models in \ref{sec:appendix}.

\section{Conclusions}\label{sec:conclusions}

In this paper we use the Millennium Simulation \citep{Springel2005} and their public catalogues to study the impact of halo mass and local density on the prediction of linear bias. 

We study the dependence of halo bias on mass (FOF mass) and local density, defined as the density fluctuation $\bar \delta$ within a given volume around each halo. Although for this paper we used a cubical box of side $l = 14\Mpc$ (which have the same volume as a sphere of radius $R = 8.68 \Mpc$), we validated that the results are similar if we use other scales. We also find similar results when we use Lagrangian instead of Eulerian densities for the background. 

 We show that bias depends much more strongly on $\bar \delta$ than on mass, and once $\bar \delta$ is fixed, the halo bias depends very weakly on mass. This is important, since it reflects that the halo bias is well constrained when the haloes are selected by the local density, almost independently on the mass. More massive haloes have a higher clustering because they are statistically in denser regions, but not because mass is the fundamental cause of clustering. In particular, low mass haloes can be found in a wide range of densities, and hence these haloes present different clustering. These results confirm the peak-background split model, that states that the large scale clustering is determined by the density fluctuations of matter, and not by halo mass. This is in contradiction with the standard HOD implementation which assumes that halo bias only depends on the mass. The local density can be seen as a property that is sensitive to other dependencies of halo bias apart from mass, such as assembly bias. Assembly bias usually refers to the difference in clustering from haloes of equal mass but different  formation time or concentration.  These differences can indeed be related to the local density, as haloes in high background densities form first and have higher concentrations. This can also explain the concept of galactic conformity \citep[see][and references there in]{Paranjape2015} by which galaxy properties, such as luminosity and colour, are not solely determine by the mass of the halo.  Thus our finding that halo local density, and not halo mass, is the key variable to understand  the large scale clustering of haloes is in line with these previous results.

To study the implication of our finding for the clustering predictions in the HOD framework, we use the method of reconstructing the linear bias from the halo bias and the occupation distribution in haloes, as explained in \citealt{Pujol2014} and in this paper.  This can be seen as a test of how well mass and local density constrain bias. More exactly, these bias reconstructions measure how well linear bias can be reproduced by assuming that the halo bias and occupation only depend on one variable (either mass or density around the halo). We show that we can predict $b_h(M)$ from $b_h(\bar \delta)$, but we cannot predict $b_h(\bar \delta)$ from $b_h(M)$. This means that $\bar \delta$ determines bias better than $M$. This is important for HOD analysis, since it is usually assumed that bias only depends on halo mass, but some galaxy populations might be affected by environment and local density as well. According to our results, the dependencies of galaxies on  the environmental density have a stronger impact on the large scale clustering than the dependencies on halo mass. With the exception of the  higher mass range, which is strongly correlated with background density, as shown in Fig. \ref{fig:mass_delta}, and therefore shows similar tendencies than local density. 

Some of the galaxy properties can be sensitive to assembly bias and the local density. In these cases, assuming that linear bias only depends on halo mass causes and error in our estimation of clustering or in the estimation of HOD parameters. Instead, we can use the local density as a proxy for bias, since it determines bias better than mass on large scales.

We show two examples of galaxy samples, $b(M_r)$ of red central galaxies and $b(g-r)$ of central galaxies, where the galaxy clustering does not depend only on halo mass. We see that the reconstruction using the local density $\bar \delta$ around the haloes makes a good prediction of galaxy bias, but the standard reconstruction using halo mass does not recover well the galaxy bias. This means, on one side, that the occupation of this population of galaxies in haloes is affected by the local density even for fixed mass, and on the other side, that even if the occupation of this galaxies in haloes depends on mass, the clustering of these galaxies is mainly due to the dependence on the local density. We also used $\bar \delta_g$ instead of $\bar \delta$ to measure the local density of haloes and the results are equivalent to those using $\bar \delta$, meaning that both $\bar \delta$ and $\bar\delta_g$ are good estimators of bias. This result is expected, since at large scales $\bar \delta_g$ is biased with respect to $\bar \delta$, but the nature of both properties is the same, and hence they disclose similar information about the local density. The advantage of using $\bar \delta_g$ is that it can be directly measured in observations, while $\bar \delta$ or halo mass are more difficult to estimate. We also show that our results and conclusions do not depend on the SAM used for the analysis. 

This analysis is focused on linear scales, where the 2-halo term dominates, so that we do not need to assume anything about the distribution of galaxies inside the haloes. A similar analysis is relevant for the 1-halo term, since small scales can also depend on local density. For example, the satellite distribution can depend on the halo concentration for fixed mass. Also, these linear scales now become more accessible with upcoming surveys. As $\bar \delta$ is a better estimator of large scale clustering than mass, and it is also easier to measure it in observations (at large scales $\bar \delta_g$ is just a biased version of $\bar \delta$), this method can be applied in observations to measure the bias as a function of $\bar \delta$, and to study galaxy clustering from the modelling of $b_h(\bar \delta)$ instead of just $b_h(M)$ to incorporate the information from assembly bias and environmental dependencies of bias.


\FloatBarrier
\section*{Acknowledgements}

We thank Ravi Sheth and Rom\'an Scoccimarro for the useful discussions about the project. We also thank Sergio Contreras for the support and discussions about the Millennium Simulation and its database. 
Funding for this project was partially provided by the Spanish Ministerio de Ciencia e Innovaci\'{o}n (MICINN), Consolider-Ingenio CSD2007- 00060, European Commission Marie Curie Initial Training Network CosmoComp (PITNGA-2009-238356). We acknowledge support from the European Commission's Framework Programme 7, through the Marie Curie International Research Staff Exchange Scheme LACEGAL (PIRSES-GA-2010-269264).
AP is supported by beca FI and 2009-SGR-1398 from Generalitat de Catalunya and project AYA2012-39620 from MICINN.
During the work on this project KH has been supported by beca FI from Generalitat de Catalunya and ESP2013-48274-C3-1-P. He also acknowledges ICTP-SAIFR, where parts of the project were done.
NJ acknowledges support from CONICET-Argentina; the European Research Council Starting Grant (SEDmorph; P.I. V. Wild); and the European Commission's Framework Programme 7, through the Marie Curie International Research Staff Exchange Scheme LACEGAL (PIRSES-GA-2010- 269264). Special thanks to Institut de Ci\`{e}ncies de l'Espai, for their hospitality.

\bibliography{biblist}

\appendix

\section{Comparison of different Semi-Analytic Models}\label{sec:appendix}

\begin{figure*}
        \centering
        \includegraphics[scale=.43]{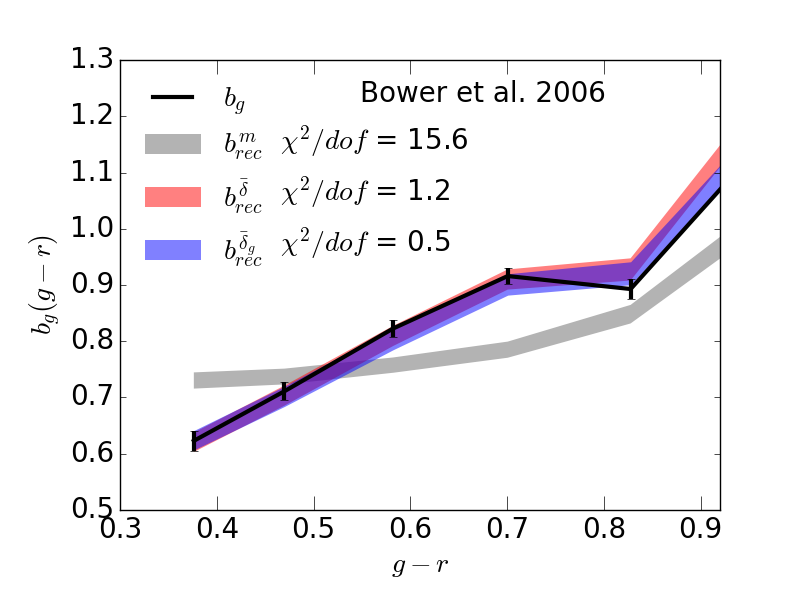}
        \includegraphics[scale=.43]{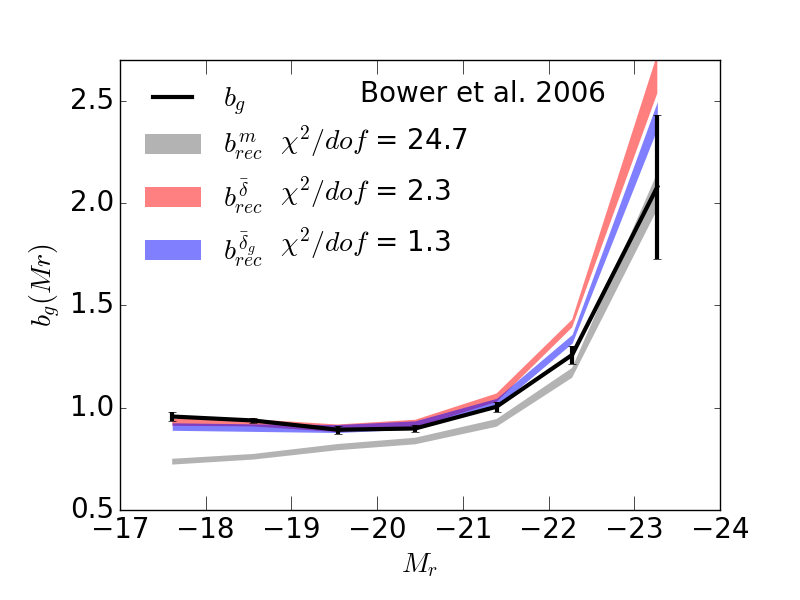}
        \includegraphics[scale=.43]{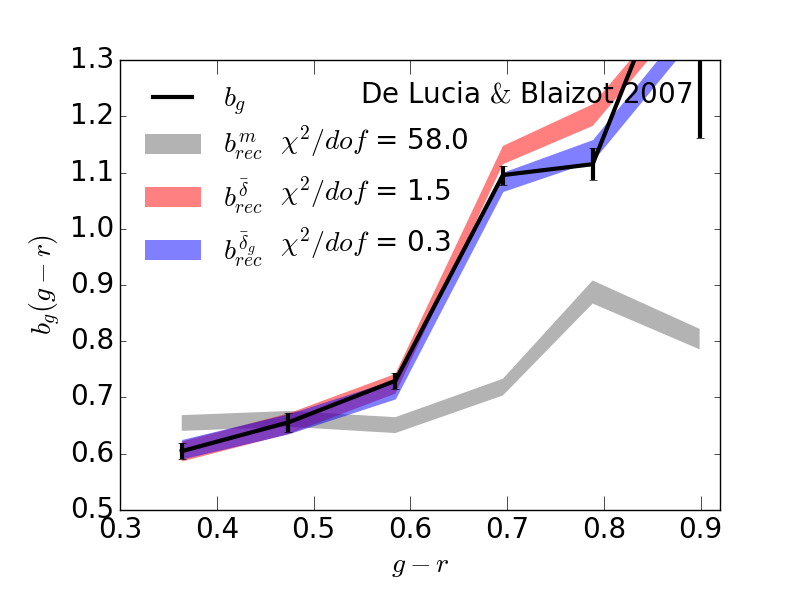}
        \includegraphics[scale=.43]{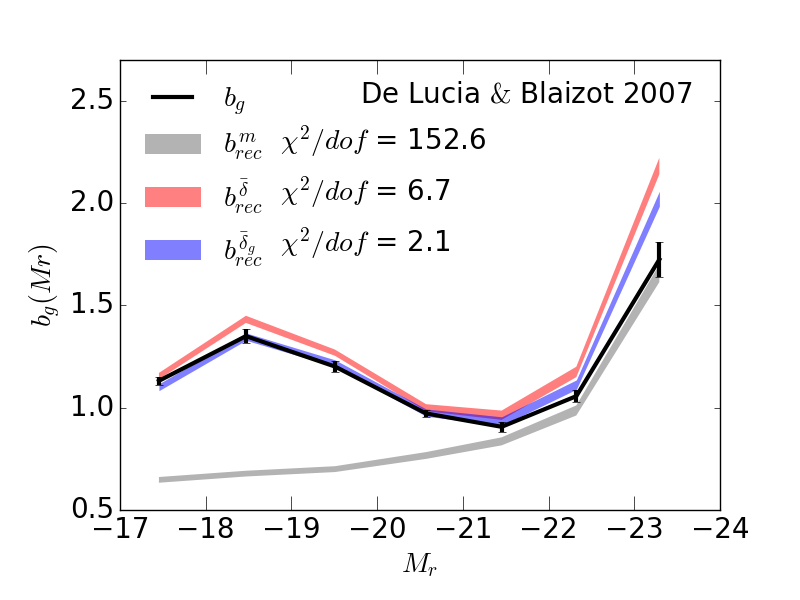}
        \includegraphics[scale=.43]{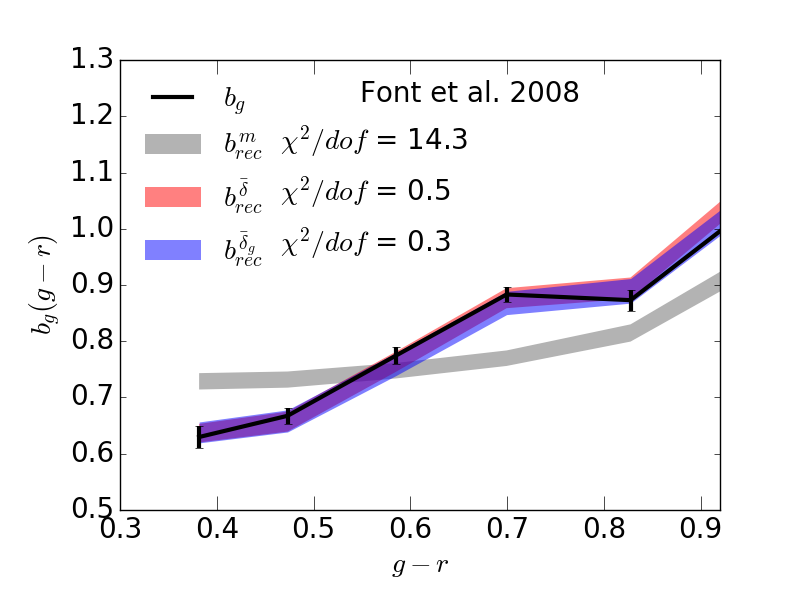}
        \includegraphics[scale=.43]{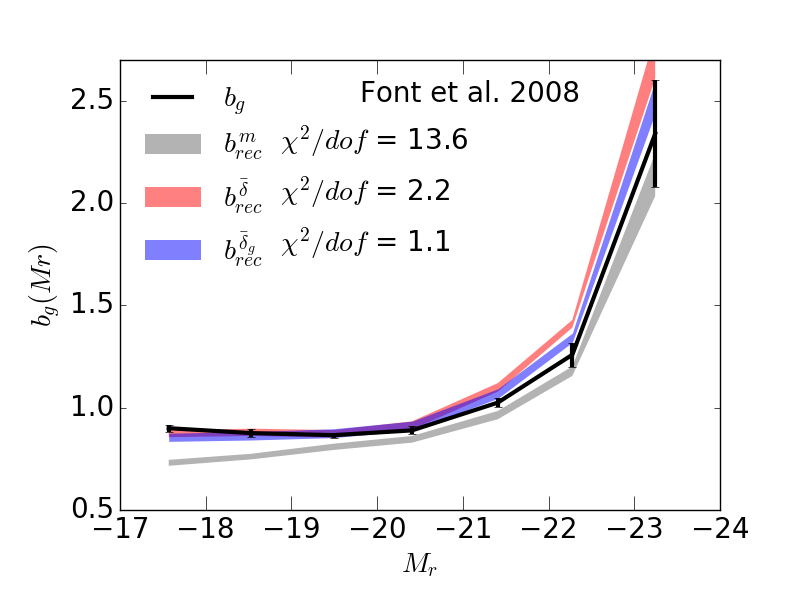}

        \caption{Comparison of bias reconstructions for different SAMs. Top panels show the results for \cite{Bower2006} model, middle panels represent \cite{DeLucia2007} model and \cite{Font2008} is shown in the bottom panels. As in Fig. \ref{fig:recs_mass}, black solid lines show the measurements of $b_g$, while $b_{rec}^m$, $b_{rec}^{\bar\delta}$ and $b_{rec}^{\bar\delta_g}$ are shown in grey, red and blue respectively. Left panels shows central galaxies as a function of colour $g-r$, while right panels show red ($g - r > 0.6 $) central galaxies as a function of $r$-band absolute luminosity.}
        \label{fig:recs_models}
\end{figure*}

In this Appendix we show the results of our analysis for three different SAMs in order to compare them with the results from Fig. \ref{fig:recs_mass} from the \cite{Guo2011} model. In Fig. \ref{fig:recs_models} we show the same results for the \cite{Bower2006} model (top panels), \cite{DeLucia2007} model (middle panels) and \cite{Font2008} model (bottom panels. As in Fig. \ref{fig:recs_mass}, the measurements of $b_g$ are shown as solid black lines, and the errors represent the $1\sigma$ level. The coloured regions show $b_{rec}^m$ (in grey), $b_{rec}^{\bar\delta}$ (in red) and $b_{rec}^{\bar\delta_g}$ (in blue) at the $1\sigma$ level. The left panels show the bias reconstructions for central galaxies as a function of $g - r$ colour, while the right panels show red central galaxies as a function of luminosity ($M_r$). The colour cut applied corresponds, as in Fig. \ref{fig:recs_mass}, to $g-r>0.6$. 

We can see that the results from \cite{DeLucia2007} and \cite{Guo2011} are very similar. The same happens between \cite{Bower2006} and \cite{Font2008}. This result can be expected because of other similarities that \cite{DeLucia2007} and \cite{Guo2011} ( and \cite{Bower2006} and \cite{Font2008}) have. In particular, \cite{DeLucia2007} and \cite{Guo2011} follow the merger trees according to the subhalo catalogues obtained from \textsc{SUBFIND} \cite{Springel2001}. On the other hand, \cite{Bower2006} and \cite{Font2008} follow the merger trees from the Dhaloes, a different definition of haloes obtained from the contribution of several \textsc{SUBFIND} subhaloes \citep{Harker2006,Merson2013}. As \cite{DeLucia2007} and \cite{Guo2011} have a common merger tree, their statistics, and in particular their clustering, are very similar (the same effect occurs for \cite{Bower2006} and \cite{Font2008}). However, the differences between the models are larger when we compare models with different merger trees. For the rest of the discussion we will refer to \cite{DeLucia2007} and \cite{Guo2011} as MPA models, and to \cite{Bower2006} and \cite{Font2008} as Durham models, according to the institutes where they were developed.

We see some interesting differences between MPA and Durham models. If we focus on the left plots, where we compare different reoconstructions of galaxy bias as a function of colour, we see that Durham models present a better agreement between $b_g$ and $b_{rec}^m$ that MPA models. This means that halo mass constraints better galaxy bias for Durham models than for MPA models. This is an example of how the agreements of the reconstructions can depend on the implementation of the galaxy formation model and the employed merger tree. Another important difference is on the right plots, where we show galaxy bias of red central galaxies as a function of luminosity. We can see that dim galaxies of MPA models present a strong clustering, while Durham models do not. The low values of $b_{rec}^m$ implies that these galaxies populate small haloes in all the models. But although the masses are similar, MPA models populate different haloes than Durham models, since the clustering is very different. In particular, red dim galaxies of MPA models populate small haloes that present a very high clustering, showing a bias of up to $1.4$. However, dim red galaxies of Durham models show a bias below $1$ in all the cases.

But although the different SAMs can present different numbers, it is important to notice that the qualitative conclusions of this study are common in all the SAMs. First of all, in all the cases mass is always the least constraining variable on galaxy bias. In all the cases, $b_{rec}^{\bar\delta}$ and $b_{rec}^{\bar\delta_g}$ show very similar results, while $b_{rec}^{\bar\delta_g}$ works slightly better. Because of this, our discussion and conclusions from Fig. \ref{fig:recs_mass} are valid for all the SAMs studied.

\end{document}